\pdfoutput=1 
\documentclass[sigconf, screen]{acmart}
\AtBeginDocument{%
  \providecommand\BibTeX{{%
    \normalfont B\kern-0.5em{\scshape i\kern-0.25em b}\kern-0.8em\TeX}}}

\setcopyright{acmcopyright}
\copyrightyear{2018}
\acmYear{2018}
\acmDOI{XXXXXXX.XXXXXXX}

\acmConference[IoTDI  '23]{}{May 9--12,
  2023}{San Antonio, TX}
\acmPrice{15.00}
\acmISBN{978-1-4503-XXXX-X/18/06}

\newif\ifrev
\revtrue
\ifrev
    \newcommand{\feng}[1]{\textcolor{orange}{[Feng: #1]}}
    \newcommand{\alan}[1]{\textcolor{purple}{[Alan: #1]}}
    \newcommand{\placeholder}[1]{\textcolor{green}{[todo: #1]}}
\else
    \newcommand{\feng}[1]{}
    \newcommand{\alan}[1]{}
    \newcommand{\placeholder}[1]{}
\fi

\usepackage{longtable}
\usepackage{tabularx}
\usepackage{tabulary}
\usepackage{geometry}
\usepackage{array}
\usepackage{lscape}
\usepackage{supertabular}
\usepackage{listings}
\begin{document}

\title{The Hitchiker's Guide to Successful Living Lab Operations}

% Alan Wang (University of Virginia) <ahw9f@virginia.edu>
% Feng Yi Change (University of Virginia) <fc4wa@virginia.edu>
% Siavash Yousefi (University of Virginia) <sy3fw@virginia.edu>
% Beatrice Li (University of Virginia) <blx2wj@virginia.edu>
% Brad Campbell (University of Virginia) <bradjc@virginia.edu>
% Arsalan Heydarian (University of Virginia) <ah6rx@virginia.edu>

\author{Alan Wang}
\email{ahw9f@virginia.edu}
% \orcid{0000-0001-6926-4336}
\affiliation{%
  \institution{University of Virginia}
  \streetaddress{Olsson Hall, Charlottesville, VA 22903}
  \city{Charlottesville}
  \state{Virginia}
  \country{USA}
  \postcode{22903}
}
\author{Feng Yi Change}
\email{fc4wa@virginia.edu}
\affiliation{%
  \institution{University of Virginia}
  \streetaddress{Olsson Hall, Charlottesville, VA 22903}
  \city{Charlottesville}
  \state{Virginia}
  \country{USA}
  \postcode{22903}
}
\author{Siavash Yousefi}
\email{sy3fw@virginia.edu}
\affiliation{%
  \institution{University of Virginia}
  \streetaddress{Olsson Hall, Charlottesville, VA 22903}
  \city{Charlottesville}
  \state{Virginia}
  \country{USA}
  \postcode{22903}
}
\author{Beatrice Li}
\email{blx2wj@virginia.edu}
\affiliation{%
  \institution{University of Virginia}
  \streetaddress{Olsson Hall, Charlottesville, VA 22903}
  \city{Charlottesville}
  \state{Virginia}
  \country{USA}
  \postcode{22903}
}
\author{Brad Campbell}
\email{bradjc@virginia.edu}
\affiliation{%
  \institution{University of Virginia}
  \streetaddress{Olsson Hall, Charlottesville, VA 22903}
  \city{Charlottesville}
  \state{Virginia}
  \country{USA}
  \postcode{22903}
}
\author{Arsalan Heydarian}
\email{ah6rx@virginia.edu}
% \orcid{0000-0001-6926-4336}
\affiliation{%
  \institution{University of Virginia}
  \streetaddress{Olsson Hall, Charlottesville, VA 22903}
  \city{Charlottesville}
  \state{Virginia}
  \country{USA}
  \postcode{22903}
}

\renewcommand{\shortauthors}{Trovato and Tobin, et al.}

\begin{abstract}
Living labs have been established across different countries to evaluate how the interaction between humans and buildings can be optimized to improve comfort, health, and energy savings. However, existing living labs can be too project-specific, not scalable, and inflexible for comparison against other labs. Furthermore, the lack of transparency in its software infrastructure inhibits opportunities for critique and reuse, reducing the platform's overall potential. In the face of climate change and global energy shortage, we envision the future of living labs to be open source and scalable to support the integration of different IoTs, subjective measures, human-building interactions, security, and privacy contexts. In this work, we share our living lab software stack and present our experience developing a platform that supports qualitative and quantitative experiments from the ground up. We propose the first open-source interoperable living lab platform for multidisciplinary smart environment research.
\end{abstract}






% \settopmatter{printfolios=true}
\maketitle

\section{Introduction and Related Works}
% \textbf{Problem statement: Living labs need to be standardized, living labs needs to support work that are not directly tied to the building, living labs need to be open-sourced
% }
% \alan{Everyone is working in their own platform, need a platform where people can work together. Interoperable platform for multidisciplinary smart environment research. Modern multidisciplinary studies have been successful in generating concepts that take into account human psychology and machine learning to achieve energy savings \cite{he2022ai}}

% We need to shape the story better. Why are IoTs in buildings to begin with? for comfort and energy and health, and cost reduction
% We are now trying to understand the relationhip between these elements (human, cost, energy, health, and other dimensions)
The average American spends more than 90\% of their lives indoors \cite{Klepeis2001,Awada2021}, and buildings account for 40\% of the total energy consumption in America \cite{eia2017much}. Together, it is unsurprising to find that if a building is properly designed and operated around the occupants' needs, preferences, and comfort levels, we can reduce consumption significantly \cite{wagner2018exploring}. In addition to reducing energy consumption, the study of the indoor environment has also been shown to have a dramatic effect on the occupant's health, and well-being \cite{Wang2022sbs,Baniassadi2021,Cedenõ2018}. These studies indicate that the improvement of the indoor environment is not only worthwhile financially but also pressing healthfully. With so many benefits, why is occupant data still so vastly under-exploited \cite{wagner2018exploring}? 

On the energy side, a lack of standardization in the production of buildings compared to the automobile industry and poor information and communications technology (ICT) infrastructure in pre-existing buildings prevents building managers from achieving the 15\%-50\% energy-saving advanced control strategies have demonstrated \cite{drgovna2020all}. On the health side, studies have shown that proper management of the environment can lead to better physiological and psychological outcomes for occupants. However, relying on employee self-reported surveys instead of quantitative measures through Health Performance Indicators (HPI) limits the potential for buildings to support occupant health and well-being \cite{allen2020healthy}. 
Researchers have created an approach called "Living Labs" to tackle these issues together. While many definitions for a living lab exist \cite{ballon2005test,niitamo2006state,bergvall2009concept}, a previous survey of existing living labs proposed a general definition \cite{cureau2022bridging}: \begin{quote}
\textit{``A living lab... is a ... typical indoor environment where everyday tasks are performed by occupants over a significant period of time to experimentally characterize their activities and responses, and with a permanent setup that allows hosting scientific experiments ... by monitoring and controlling the indoor conditions..."}
\end{quote}

% \textit{``A living lab for human-centered comfort investigations indoors is a whole building or any space designed as a typical indoor environment where everyday tasks are performed by occupants over a significant period of time to experimentally characterize their activities and responses, and with a permanent setup that allows hosting scientific experiments about environmental comfort by monitoring and controlling the indoor conditions and/or the space layout."}

However, the definition assumes certain qualifiers that make it flexible for interpretation. For instance, what is a \textit{typical indoor environment}, what are \textit{everyday tasks}, and what is \textit{a significant period of time}? We sampled a list of living labs from the surveys and extended columns about survey distribution, device deployment, and accessibility in Table \ref{tab:lls}. Notably, most existing living labs do not have their software infrastructure code readily accessible. We identify this lack of a generalizable ICT infrastructure as an opportunity to promote international collaboration and to retrofit existing buildings with a platform to design and test smart indoor environment applications. Research demonstrating this bottom-up approach to smart environments can already be seen in innovations for rooms \cite{zhao2017mediated} and through smart furniture applications \cite{aryal2019smart}. These studies not only enable more user-centered control schemes to target environments at a micro-climate level but also give researchers the freedom to explore interactions that, if failed, will not compromise essential building systems. 

\begin{table*}[!htp]
    \centering
    \begin{tabular}{>{\raggedright}p{1.5in}>{\raggedright}p{1in}p{0.5in}p{0.5in}p{0.5in}p{1.25in}p{0.5in}}
    \hline
    \textbf{Name} & \textbf{Country} & \textbf{Source Code Accessible} & \textbf{Surveys Distributed} & \textbf{Sensors Deployed}  & \textbf{Longest Study Duration} & \textbf{Reference} \\
    \hline
    BNZEB & Italy & N & N & Y & 1+ years & \cite{ascione2020nearly}\\
    && & N & Y & 1+ years & \cite{ascione2019framework}\\
    \hline
    Carleton University Buildings & Canada & N & N & Y & 15 months & \cite{bursill2020multi} \\
    & & & N & Y & 7-12 months & \cite{gunay2018development}\\
    & & & N & Y & 1 month preliminary data collection + 9 weeks study& \cite{gilani2018preliminary}\\
    \hline
    CIRIAF Offices & Italy & N & Y & Y & 1 year & \cite{pisello2016peers}\\
    & & N & Y & Y & 1 year and 2 years & \cite{piselli2019occupant}\\
    \hline
    David Bower Center & California, USA & N & Y & Y & 2 months & \cite{roa2020targeted}\\
    \hline
    ENERGISE Living Labs & Denmark, Finland, Germany, Hungary, Ireland, Switzerland, the Netherlands, and United Kingdom & N & Y & N & 11 months & \cite{sahakian2021challenging}\\
    \hline
    House Living Labs Project & Australia & N & N & Y & 1 year& \cite{eon2018influence}\\
    \hline
    Sutardja Dai Hall at University of California, Berkeley & California, USA & N & N & Y & Two weekend days &  \cite{ma2011optimal}\\
    \hline
    Well Living Lab & Minnesota, USA & N & Y & Y & 18 weeks & \cite{clements2019spatial}\\\textbf{}
    & & & Y & Y & 4 weeks & \cite{zhang2020impacts}\\
    & & & Y & Y & 14 weeks & \cite{jamrozik2019access}\\
    \hline
    ZEB Living Lab & Norway & N & N & Y & N/A (description of the test facility) & \cite{goia2015zeb} \\
    \hline
    \end{tabular}
    \caption{Sample of living labs constructed after 2000 (adopted from \cite{cureau2022bridging}). Studies can last anywhere from 4 weeks to 2 years, almost all the labs have sensors deployed, and some use surveys as an additional input stream. We could not readily access the underlying infrastructure for any of these labs.}
    \label{tab:lls}
\end{table*}

While applications often receive more direct attention from occupants, the infrastructure and platform layers are also significant areas to address \cite{dryjanski2020adoption}. In this work, we elect to focus our contributions in the \textit{living lab ICT software infrastructure} domain. Specifically, we share the lessons we learned while attempting to create a living lab from the ground up. We then propose a set of standardized components for living labs infrastructure and open source our code to support future research teams in streamlining their software development efforts.

We organize our paper as follows: we first describe the problems we've encountered while setting up a living lab (Section \ref{sec:problem}). Then, we elaborate on the solution by describing the overview of our framework and the relationship between the problems and the components (Section \ref{sec:solution}). We then describe the solutions and lessons we've learned when establishing a living lab (Section \ref{sec:lessons}). Then, we point towards limitations and directions for future work (Section \ref{sec:limits}). Finally, we give our concluding remarks (Section \ref{sec:conclusion}), acknowledgements (Section \ref{sec:ack}), and share the online resources (Section \ref{sec:online}).

\section{Challenges}\label{sec:problem}

In this section, we share the problems we encountered as we set up our living lab from the ground up, leading to the architecture and framework described in Section \ref{sec:solution}. 

\subsection{Sensor, participants, and surveys can scale faster than the research team}\label{sec:sensor_fail}

One of the first challenges we encountered when setting up our living lab was the registration of newly arrived devices and their locations. At first, we settled the problem by having a building information model (BIM) and an excel sheet with some device metadata. However, as the number of participants and sensors grew, we realized that it was ineffective to manually update the BIM model or the excel sheet whenever researchers introduced a new device or an occupant submitted a maintenance request. We identify a need to \textit{enable granular control of access to smart environment metadata information}, including keeping track of an inventory of cyber and physical assets. 

Keeping this inventory is helpful because, over time, the same issue for managing the physical object of a living lab can manifest itself in tracking the system's data quality. For example, one of the more insidious problems we encountered came from sensors that were still streaming data. However, information became lost in transit due to the number of walls between the sensor and the gateway receiving the data \cite{wang2020my}. If we had assumed all sensors that were streaming data to be valid, we would have missed an increasing amount of data loss over time. Another problem we encountered was modern-day ``smart-outlets", which disabled themselves based on their local occupancy sensor readings. In other words, data continued to stream until late at night when no occupants were at the lab and when the researchers might be sleeping. If we had a dashboard platform where we could visualize the system's activity at night, we would have been able to observe this issue. Finally, we also encountered the challenge of human hazard, as previously reported \cite{hnat2011hitchhiker, barrenetxea2008hitchhiker}. The daily activity of the occupants (e.g., running into walls) dislodged sensors and removed gateways. Through all these challenges, we identified a need to \textit{locate invalid data through dashboards and automated routines}. Visualizing plots and automatically discovering outliers reduce the maintenance burden for the research team and enable consistent data collection.

\subsection{Proprietary, heterogeneous software, hardware, and skill sets can limit the ability for a team to work together} 
During our experience collaborating with other labs, teams generally enter with a different collection of physical and digital tools, bolstering the number of interactions and avenues for research. However, many of these IoTs come with companion algorithms and online platforms, leading to avoidable subscription fees and repeated development efforts to integrate the tools. Furthermore, keeping a chain of custody for these different software and hardware components becomes burdensome and difficult to assign among the labs and participants. These challenges exponentiate when a mixture of collaborators with different skill sets, backgrounds, and interests come together. We either needed to build additional infrastructure to improve access or ran into bottlenecks for the operation that rose to the combined level required to navigate the tools, often leading to projects with a low truck factor \cite{avelino2016novel}. Consequently, we identified a need for\textit{a living lab platform that can reduce the technical barriers of entry and enable people to maintain it with a variety of skill sets}.

While lowering the barrier to entry enables different people to help with a living lab operation, to do research involves exploring new areas and implementing connections that might not yet exist. We realized that living lab platforms mandate the ability to integrate new research areas into an existing ecosystem of people and devices. For instance, using self-powered sensors can require a gateway topology, requiring researchers to consolidate between different time scales and radio protocols. We also were allowed to integrate autonomous sensors (i.e., robots), which required the ability to fit the robot-sampled environmental variables into a shared spatial coordinate with other known devices \cite{jin2017indoor}. Finally, edge computing paradigms challenge assumptions of deployment locations for living lab platforms. For example, IoTs can sometimes better be deployed on the edge without utilizing cloud platforms \cite{nasir2021enabling}. Through this large variety of topics, we identified a need \textit{for a living lab platform to enable the interaction of heterogeneous devices, software, and changing paradigms of computing}.

\section{Framework Description}\label{sec:solution}
In this section, we describe the framework we have built to address the needs identified in the previous section. We first describe an overall architecture of the system (Section \ref{subsec:framework_overview}). Then, we elaborate on the individual modules and considerations (Section \ref{subsec:design_decisions}). 

\subsection{Overview}\label{subsec:framework_overview}
\begin{figure}[!h]
  \centering
  \includegraphics[width=\linewidth]{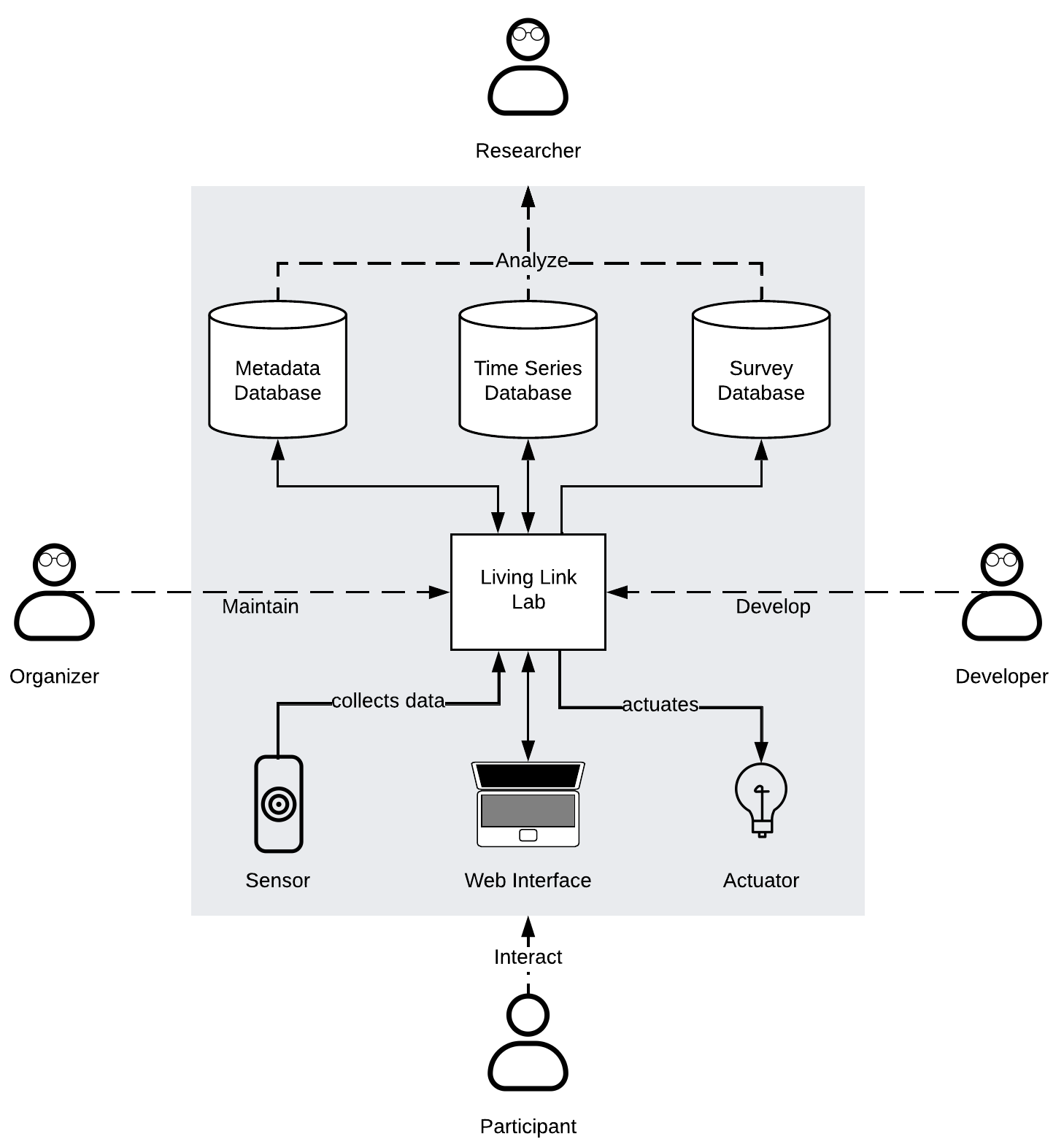}
  \caption{Physical overview of the Living Link Lab System. The bottom-up architecture assumes that people can start their research agnostic to existing building infrastructure systems.}
  \label{fig:lll-big}
\end{figure}

Figure \ref{fig:lll-big} showcases a conceptual diagram of the living lab system. We separate digital and physical types. Physical members are researchers, developers, organizers, and participants. A researcher analyzes and interprets the collected information from the environment and the participants. Participants are occupants who are enrolled in a living lab study. A developer focuses on updating and maintaining the software infrastructure, and an organizer helps with the operations from a limited-technical capacity. The digital representation specifies an increasing amount of permissions, from users to staff, to administrators. For example, a user has access to the web interface but cannot access privileged views that a staff user can. Staff users can additionally gain permission to modify the value of models stored in the framework. An administrator privilege gives a user staff privileges but additional the ability to create or delete models. In other words, a user is permission granted to people in cyberspace with access to the system, which stands distinct from an occupant, anyone who physically occupies the building. People who dwell in the building but do not subscribe to the system are considered occupants but not users (i.e., a non-participant). Researchers can be a user and not an occupant. We make these conceptual distinctions because we recognize during our deployment that blanket assumptions about the technical skill set and categorical designation for people in the building can limit the research teams' ability to interact with the community. We expect future users of living labs will face similar difficulties, so summarize these distinctions in Table \ref{tab:ll_users}.

\begin{table}[]
    \centering
    \begin{tabular}{p{0.7in}p{0.5in}p{0.5in}p{1.2in}}
        \hline
        \textbf{Physical Descriptor} & \textbf{Technical} & \textbf{Cyber Permissions} & \textbf{Role}\\
        \hline
        Researcher & No & Admin & Oversees the operation of the living lab and analysis of data\\
        \hline
        Developer & Yes & Admin & Maintains and develops the framework\\
        \hline
        Organizer & No & Staff & Maintains and develops the relationship between participants and researchers, and also support with the logistics the research study\\
        \hline
        Participant & No & User & A person who interacts with and has data collected to the living lab\\
        \hline
        Non-participant & No & Non-user & A person who occupies the building that a living lab study is being conducted\\ 
        \hline
    \end{tabular}
    \caption{Conceptual organization of members. Limiting permissions for different types protect the system from invalid use and also effectively consolidates responsibilities. The technical column explains the necessity for the research team member to know how to programmatically engage the system (instead of engaging a Graphical User Interface (GUI)). }
    \label{tab:ll_users}
\end{table}

    % \item \textbf{Administrators}: Non-technical users who are concerned with the relationship between participants and the research, who also support with the logistics of user, device, and surveys
    % \item \textbf{Developers}: Technical users who focus on the maintenance and development of the web framework 
    % \item \textbf{Participants}: Technical and non-technical users who interact with the framework

\begin{figure}
    \centering
    \includegraphics[width=\linewidth]{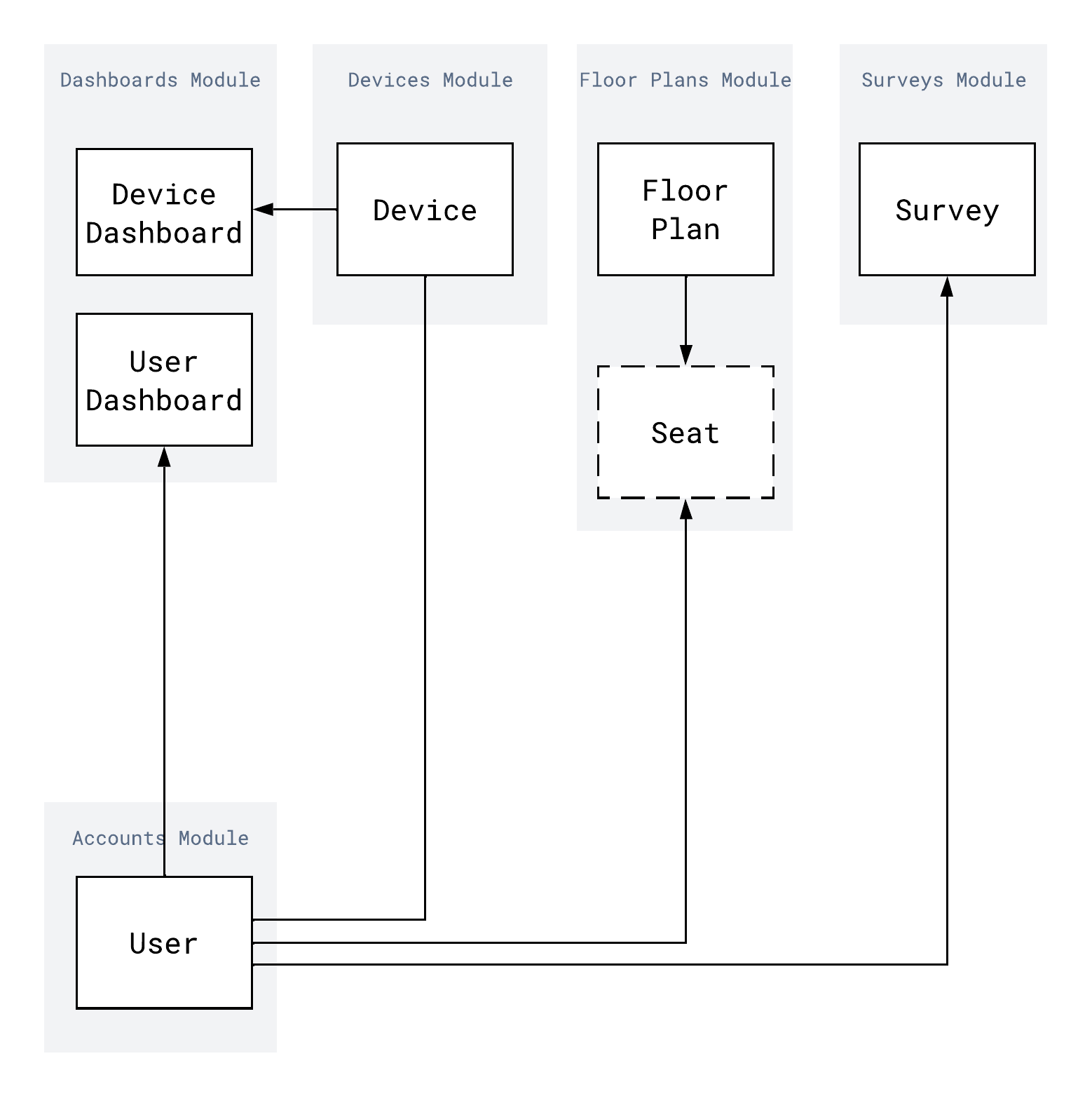}
    \caption{Cyber model diagram describing the dependency relationship for occupants in living labs. The direction of the arrow describes the sequence with which model objects are deleted when they are removed. For example, deleting a user from the system triggers a cascade of deletions downstream, removing the user's dashboard, the user's seating information, and all associated survey metadata. However, removing a custom user will not delete registered devices from the system, nor delete floor plans or surveys.}
    \label{fig:lll_high-level}
\end{figure}

In addition to the conceptual categorization of use cases, the implemented living link lab web framework uses a Model View Template (MVT) architecture \cite{vincent2021django}. Specifically, a model provides an interface to data stored in a database, a template handles all static components of a web page, and a view renders a response to the user by combining information drawn from models and templates. Generally, under each module is a testing sequence and a view that takes the model and template and serves an HTTP response to the participant. Figure \ref{fig:lll_high-level} shows a high-level relationship graph between the different modules. For example, a participant can be digitally represented and have a connection with a set of devices, locations, and surveys. First, a researcher creates a floor plan representing the location where occupants can inhabit. Then, a participant is assigned a seat, relating the participant to the floor plan. The system can then use the distance between participant seating and registered device locations to assign specific sensors to the participant. When a participant is created and assigned a sensor, the system creates a participant-specific dashboard and a device-specific dashboard. Finally, a digital representation is created in the survey model when a participant is assigned surveys. For further details about each of the models and fields, we include a detailed export of our existing system in Figure \ref{fig:lll_model_detailed} of the appendix.

% Compared to traditional roles where occupants might only have control over their own desk lamp, we envision a system that \textit{can} coalesce the preferences of individual participants and calculate an optimal control scheme to tailor to the collective environment, looking towards granting users and buildings more ways to interact \cite{sutherland1965ultimate}.

\subsection{Design Decisions} \label{subsec:design_decisions}
In order to build in extensibility while minimizing complexity, we operated under the assumption that each module should be only loosely-coupled with the user's digital representation (the user's \textit{model}). Specifically, the less code required to sustain a system, the easier it can be to debug and maintain it. For example, we anticipate that some use cases might not need a survey module (as shown in Table \ref{tab:lls}), so the framework allows researchers to remove the surveys module in settings.

We separated the components that would benefit from a living lab infrastructure into the following modules: \textit{accounts}, \textit{surveys}, \textit{devices}, \textit{floor plans}, and \textit{dashboards}. The \textit{accounts} module encapsulates user metadata and access. The \textit{surveys} module encapsulates the survey metadata. The \textit{devices} module encapsulates the device metadata. The \textit{floor plans} module stores different environmental contexts which allow tagging of user or device into the time series database, and the \textit{dashboard} module automates the generation of panels based on created devices. Below, we elaborate on the roles and functions of these modules.

\paragraph{Accounts}
The accounts module represents the user in digital form, which contains the addition of necessary information to the participant's metadata, such as age group or occupation. For example, an organizer with staff user privileges can read compliance views and write to the \textit{surveys} model to help handle the day-to-day operation of surveys. For example, if a user needs more time to complete a survey or if a user does not receive an email, an organizer would be able to edit the system to redistribute or extend the deadline for surveys. However, limiting the privileges disallow organizers to create or delete \textit{accounts} model. This lack of admin user privileges means that staff users cannot inadvertently delete a researcher's account when they edit the system. 

The user authentication workflow generally works ``out of the box" using modern web frameworks such as Django, which also helps alleviate the burden on the researchers to maintain the most up-to-date security features (something commonly overlooked by living lab researchers and organizers). In our framework, we utilize three classes of privileges as described in Table \ref{tab:ll_users}: 1) \textit{user}, who interacts with the system with no additional privileges, 2) \textit{staff}, who can gain read and write and gain access to restricted views such as a survey compliance dashboard or specific admin data as assigned, and 3) \textit{admin}, who has the highest level of access to the web interface of the system, who can not only alter values for existing models but can also create and delete objects to improve the operation of the framework. For instance, admin users have the authority to remove staff users or delete other users from the system. These distinctions help facilitate the smooth scaling of the living lab operations by allowing researchers to share responsibility. 

\paragraph{Floor Plans} \label{par:floor-plans}
We include a module called \textit{Floor Plans} because we found it essential for users to note their seating arrangements during our deployment. Additionally, modern-day occupancy habits no longer conform to a single permanent seating location. Hybrid work environments and hoteling spaces make tracking multiple spaces for users mandatory in certain situations. We couple the floor plans model with the user and device models, allowing us to explore custom ways to automate device visibility with users. For example, a custom routine can be created to assign all known devices of a building to users within a fixed geometric radius to their assigned seating. Floor plans are critical because it enables the adding, removing, and relation of digital representations of physical locations.

\begin{figure}
    \centering
    \includegraphics[width=\linewidth]{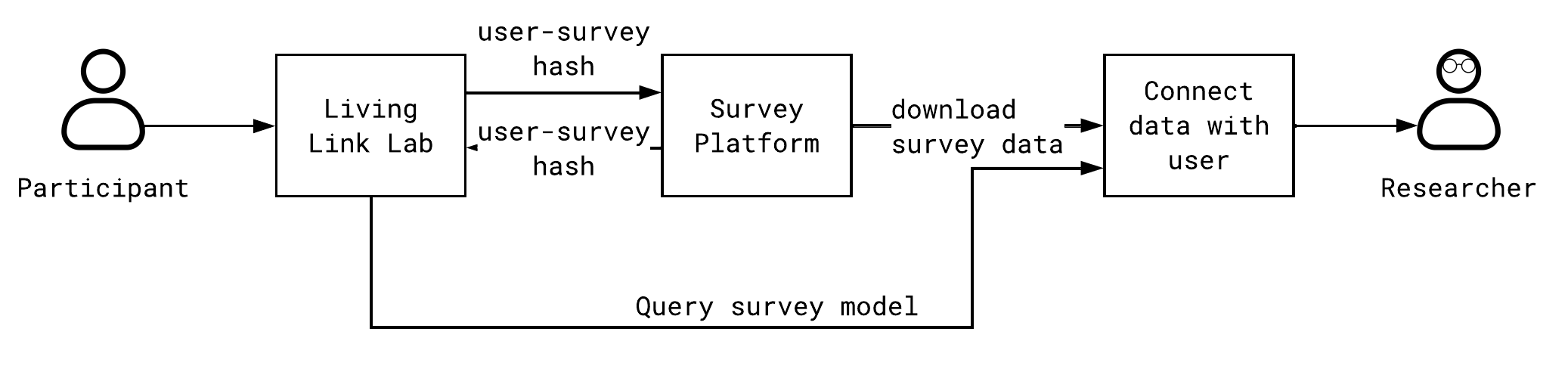}
    \caption{Example survey workflow: participants can be protected from identification from researchers downstream who only have access to the survey data if the survey identifier is a hash of user-survey specific information. A researcher would require both the user and survey metadata information to retrieve the hash to identify the survey owner.}
    \label{fig:survey_pipeline}
\end{figure}

\paragraph{Surveys}
We created a surveys module because we found that designing surveys correctly often requires a dedicated service such as Qualtrics. The survey system must be flexible to support the distribution of surveys based on the research or study's needs. Based on our experience and reviewing previous living lab studies, surveys could be distributed at the occurrence of an event, daily, weekly, monthly, or other set periods. Furthermore, tracking survey compliance is another element from which the organizer and researcher can benefit. Such compliance checks can be done internally to enable real-time adjustments and delivery strategy testing. As a result, we implemented an "anonymous link" survey workflow, shown in Figure \ref{fig:survey_pipeline}, where we keep track of the association between user-and data by altering the ensuing hash of user information. For example, if we hash the user password, researchers with only access to survey data cannot identify the user who took the survey. We can also create an identifier by including survey metadata and user information, resulting in unique hashes for every row of survey data based on user-survey pairs. By doing so, a researcher can only identify which participant answered the survey by accessing the metadata and survey databases. 

\paragraph{Devices}
The devices module contains critical pieces of information to help streamline data upload and retrieval. Specifically, we store the ``tag" information for time-invariant data and open APIs to stream ``field" information for time-variant data. For example, we can include a URL for adding humidity as an input stream to a device or a URL to check and only upload data whose field matches known fields. We include an example data point in Figure \ref{fig:example_dashboard} to help illustrate the process and the concept between fields and tags.

\begin{figure}
    \centering
% \begin{minted}[fontsize=\small, frame=single,framesep=10pt,tabsize=2,breaklines]{python}
% {
%     'time': '2020-12-23T23:54:50.727Z',
%     'device_id': '503eaa71b92a', # tag
%     'location_general': 'Link Lab', #tag
%     'location_specific': 'grid_5', #tag
%     'fieldname': 'heartbeat' # tag
%     'system_version': 'lll-1.0.0', #tag
%     'value': 1, # field
%     'counter': 256, #field
% }
% \end{minted} 
    \begin{lstlisting}
        {
            'time': '2020-12-23T23:54:50.727Z',
            'device_id': '503eaa71b92a', # tag
            'location_general': 'Link Lab', #tag
            'location_specific': 'grid_5', #tag
            'fieldname': 'heartbeat' # tag
            'system_version': 'lll-1.0.0', #tag
            'value': 1, # field
            'counter': 256, #field
        }
    \end{lstlisting}
    \caption{Example data point written to the time series database. Generally, \textit{tags} are variables to query on that are more time-invariant, while \textit{fields} are variables that change more dynamically.}
    \label{fig:example_datapoint}
\end{figure}

In the figure, \textit{time} tracks when the data point is uploaded. The \textit{device identifier} is a unique value that identifies the devices, \textit{location general} allows for context-specific coordinates that are specified in \textit{location specific}. For example, \textit{location general} can be the building where the device is placed, and \textit{location specific} the room the device is in. The \textit{field} variables enable visualization of different devices onto the same dashboard, such as humidity. The \textit{value} stores the sampling information. For instance, humidity can be a \textit{field}, and 40\% can be a \textit{value}. However, we only track the device identifier, the location general, and the location specific to the metadata database. For the actual data entry, we enable API access on the website to pipe information directly into a back-end time series database. It is important to note that there is a difference between \textit{tags} and \textit{fields}. For our database, the difference is that one signifies which variables will be queried versus which variables will just be tracked.

\paragraph{Dashboards}
% Grafana is an open-source observability platform that pairs well with influxDB. 
During our deployment, we found that having real-time dashboard users can access was a greatly desired feature. Real-time dashboards enable users to monitor their surroundings and potentially enable new avenues for users to utilize their microclimate information. However, we found that dashboards also have a public and private components. Some public information is irrelevant to the user, such as the air quality of the building next door that they do not care about sharing. However, some information is private, such as the occupancy schedule detected at their seat, which they mind sharing with others. To adjust to the needs and ethics of information distribution, we approached the design of the dashboard module by separating a user-specific dashboard and a device-specific dashboard. User dashboards allow users to customize the information relevant to their day-to-day. In contrast, device dashboards contain information about the individual device, which enables device-specific insights and debugging. We make no stipulations about which information should or should not be shared. We mainly provide opportunities for developers and researchers the space to pick and choose how the assignment policy should be. 

\section{Lessons learned}\label{sec:lessons}
In this section, we describe our lessons learned and the requirements we have observed that map to one or a combination of the module described in the previous section.

\subsection{Quickly see if a sensor is working}
\begin{figure}
    \centering
    \includegraphics[width=\linewidth]{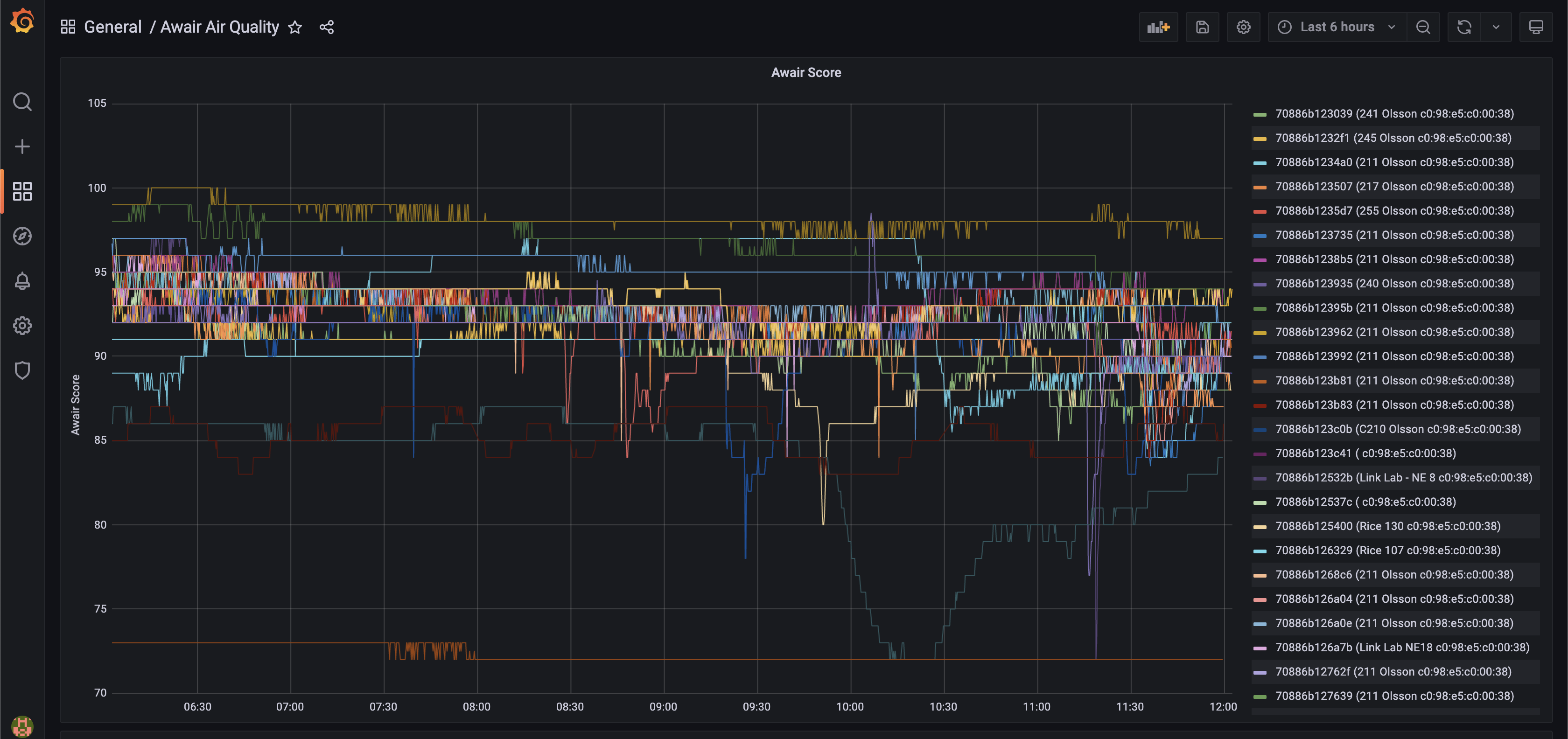}
    \caption{A generated dashboards showing real-time air quality data for the last 6 hours.}
    \label{fig:example_dashboard}
\end{figure}

When installing a sensor, reducing the number of steps to generate a corresponding dashboard of the sensor's collected values helps researchers validate the sensor. In most cases, owners of devices only want to verify that data is actually streaming in from the devices and does not need to be replaced. For example, a light sensor can be observing data within range of its data specification sheet, but showing zero lux when light is on and 1,000 lux when light is off. A casual inspection of the signals with software will not be able to find this difference, but a researcher could catch this error to mark and fix the faulty device. However, having a dashboard is also tremendously rewarding when everything is functioning correctly as the researcher can see a possible noteworthy trend or issues with the sensor. In Figure \ref{fig:example_dashboard}, we showcase an example of a generated dashboard that organizers can create without coding.

\begin{figure}[!htp]
    \centering
    \includegraphics[width=\linewidth]{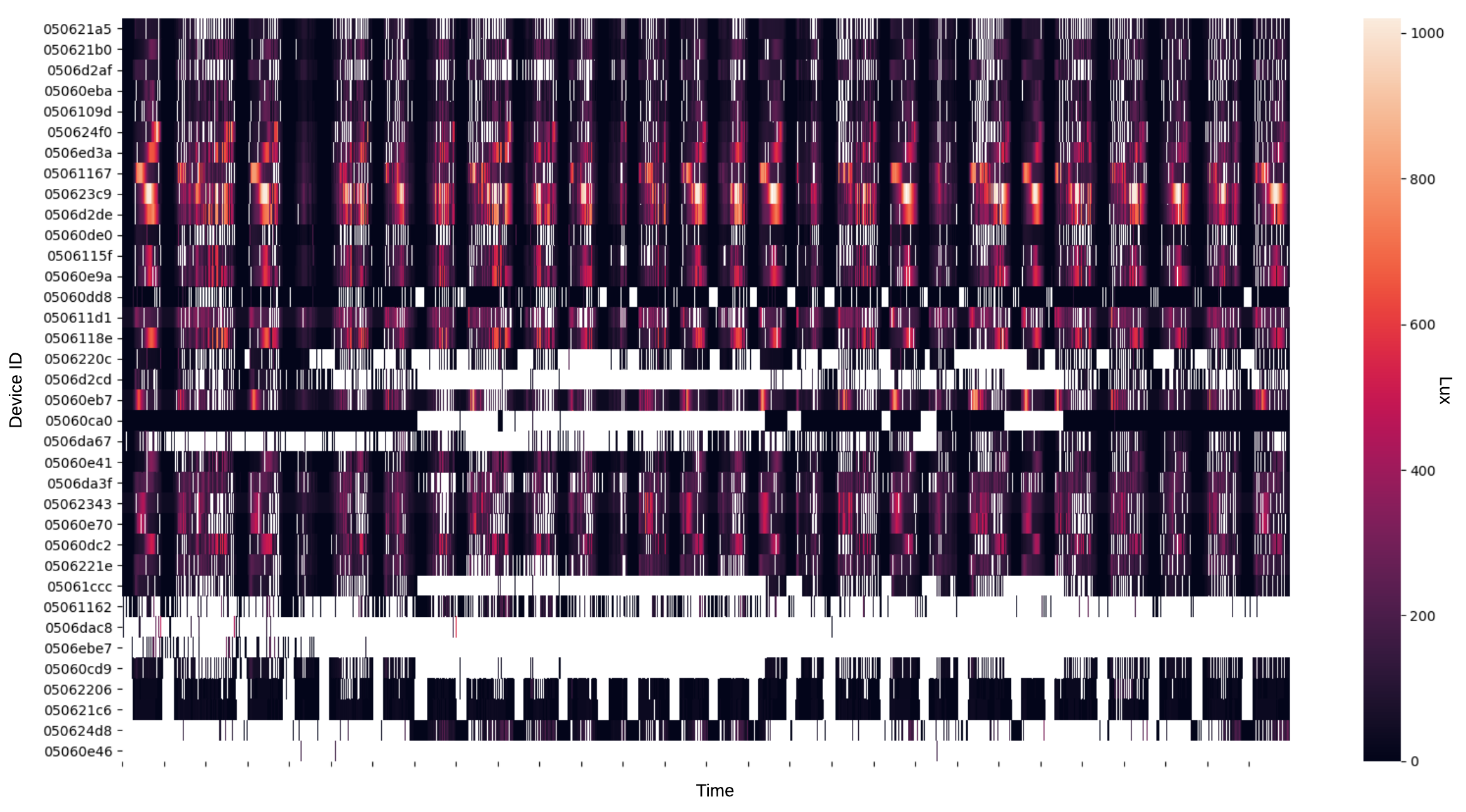}
    \caption{A month of lighting data, showcasing partial and complete data loss for more than 30 sensors.}
    \label{fig:light_month}
\end{figure}

In addition to manual inspection, automated fault detection can assist researchers in discovering more insidious faults. In figure \ref{fig:light_month}, for example, we show that sensors can fail less noticeably by collecting partial data and ways to detect it \cite{wang2020my}. Having the device model paired with the floor plans module allows us to quickly identify which sensors are malfunctioning and quickly fix or replace them if needed. However, as we have described Section \ref{sec:sensor_fail}, having a system that does not need to sleep routinely check in place of the user would free the researcher up to do other developments that can not easily be automated, such as installing the sensor or onboarding participants. Without knowing where the sensors are, we cannot quickly locate them to replace them. We did realize that it might be challenging to map precisely the coordinate of where the device is concerning a pre-defined origin in space; therefore, we introduced the grid system so the users can change the resolution to a level they can support. Consensus-based methods can also be used to help flag outliers in collected data \cite{chen2018trust}.

\begin{figure*}
    \centering
    \includegraphics[width=\linewidth]{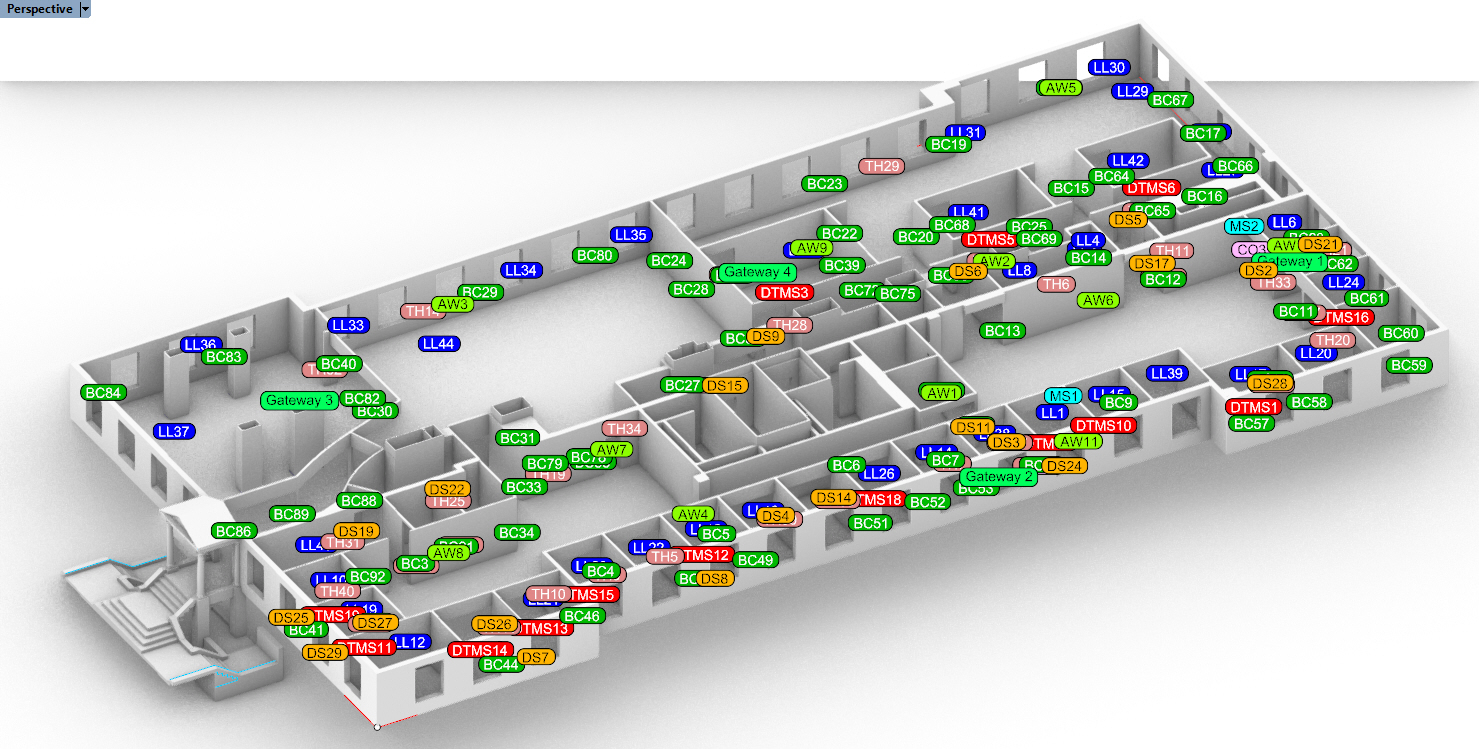}
    \caption{Tracking device locations in 3D software enable researchers to record precise locations and spatial context. However, the skill set required to model and manipulate the digital model makes labeling and recording the location of devices prohibitive for organizers without modeling skills. Furthermore, licensing or other operating systems requirements make it challenging to access and maintain automated routines to track changes across time.}
    \label{fig:lll_device_plan}
\end{figure*}
 
\subsection{Allow for flexible survey delivery}

% \begin{itemize}
%     \item Assume that users will be on-boarded in a rolling basis
%     \item Make routine surveys as simple and short as possible
%     \item Enable multiple avenues to access surveys if available (email, calendar invites, text message)
% \end{itemize}
During our framework deployment, we realized that we often ran into issues about how the surveys were deployed, such as needing to re-word and remove questions that might have been redundant. Furthermore, we ran into situations where we had to onboard participants at separate times of their convenience, which meant that we could not do things like send out a mass email after everyone had been registered for the study. These challenges come in addition to needing to calculate and see which surveys have been completed to reimburse the participants correctly. 

The survey module is designed with open times and close times and pairs with users. We marked open time, user, and survey URLs as unique together because, individually, they can repeat with survey objects. For example, different users can be subject to the same surveys simultaneously at the same time. Having user-survey-time-specific objects also allows the organizers to automate portions of survey emailing and track compliance in real time. For example, we show a quick example in Listing \ref{code:user-survey-query} where developers can quickly filter surveys by user and by completion status.

\begin{figure}
    \centering
    \begin{lstlisting}
        from surveys.models import Survey
        
        total_user_surveys= Survey.objects.filter(user__username = <username>)
        total_user_surveys_completed = total_surveys.filter(completed=True)
    \end{lstlisting} \caption{listing}{Query number of completed surveys for any given user in real-time}
    \label{code:user-survey-query}
\end{figure}

\subsection{Allow users and devices to move}
One of our most rewarding experiences came from the need to deploy a Temi-robot \footnote{https://www.robotemi.com/} to sample environmental components. We quickly realized a need to make a mental model for which aspects of our system are dynamic and which parts of the system are static. For example, static components can be stored as \textit{tags}, which enables us to query the database and find relevant data about that item. This manifests in things like device identifiers or the floor where the sensor is installed. However, for a robot, what could have been a tag can now become a \textit{field}, or things we need to record but do not make queries directly. For example, tracking the robot's x, y, z coordinates through space with floating points might not reoccur for long periods, reducing the value gained with the query. In this example, though, we demonstrate a fundamental tension between the need to track movements in space and the immutable nature of the environment. Towards this end, we implemented grids, making it easier for users to query and record sensor locations but giving room for a more granular \textit{field} such as ``coordinates".

\subsection{Minimize the technical skill set required for researchers and organizers to participate in the operation of the living lab} 
Depending on the team size and resources, it may be infeasible to expect everyone in the research group to have the necessary software development skills. However, checking for user compliance and reaching out to users are essential tasks that can be accomplished without technical skills. By incorporating user interfaces that allow for the query and modification of data without needing to write code, developers can empower organizers to help maintain and fix the operations of the living lab. Additionally, incorporating traditional web forms as opposed to navigating computer-aided design software such as Rhino in Figure \ref{fig:lll_device_plan} reduces the technical barrier for organizers and researchers to help track sensor location information. 

\subsection{Allow participants to control and protect their data}
Collecting long-term sensitive participant information can quickly become privacy-intrusive. Creating schemes to enable users to have the pro-activity to delete data and observe their information gives control and ownership of information back to the participants. Exposing models into forms and APIs such as through Django Rest Framework \footnote{https://www.django-rest-framework.org/api-guide/schemas/} enables participants to make changes to the model controlling the system. Similarly, using data coupling instead of control coupling \cite{jasinski2016effective} between the web framework and associated applications allows the severing of the relationship without deleting the user.

\begin{figure}
    \centering
    \includegraphics[width=\linewidth]{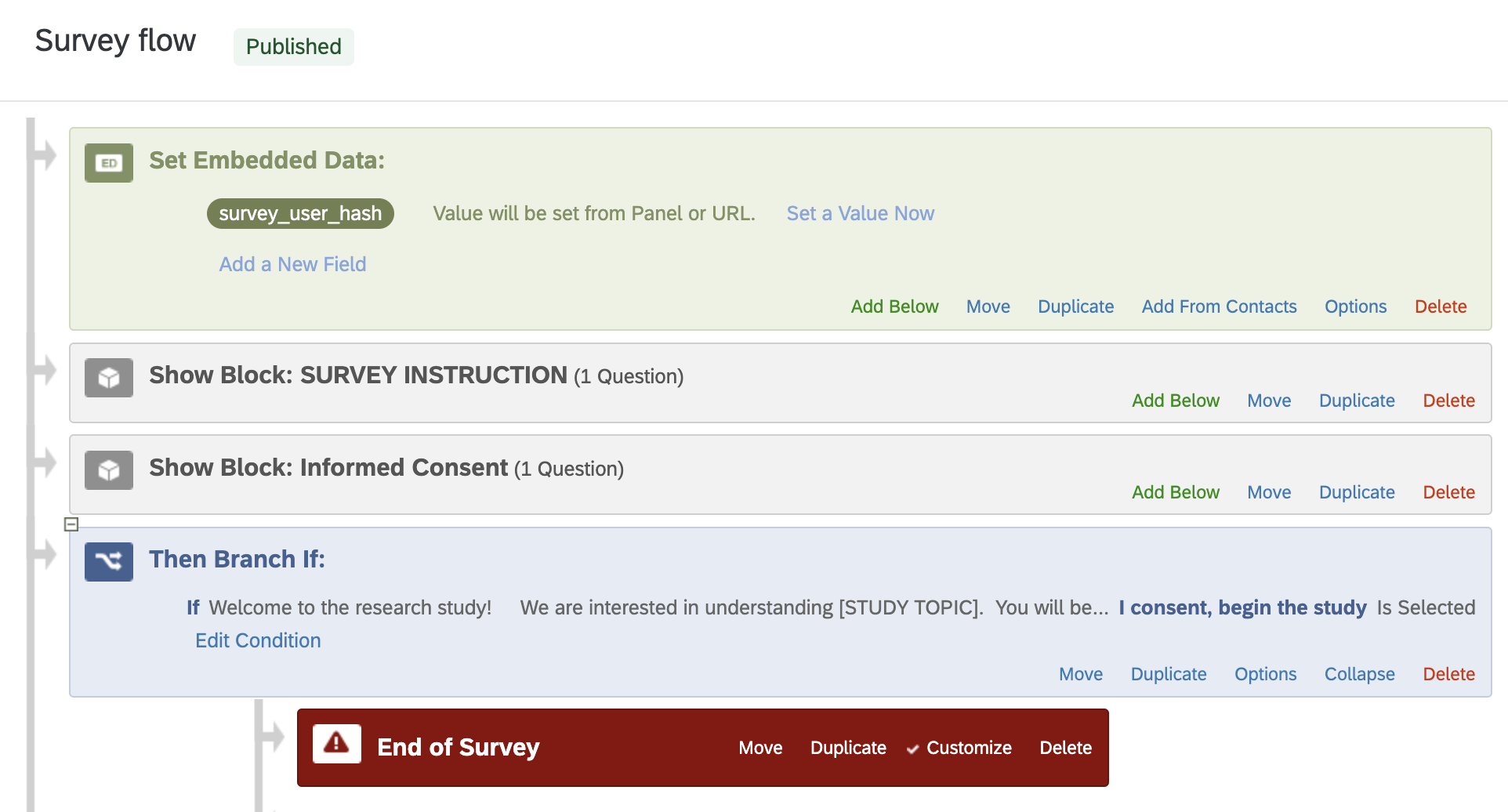}
    \caption{Example screenshot from Qualtrics. Embedding a hash of the user information alongside with the surveys identifiers allow participants to remove themselves from prior identification by changing the portions of the hashed information (e.g., changing the password changes the user-identifier on the survey platform).}
    \label{fig:survey}
\end{figure}

\subsection{Connect to humans as soon as possible and beware the timescales}
We observed interesting environmental trends across annual periods during our living lab infrastructure setup. However, we could not make any claims about the relationships between health aspects and environmental exposure without connecting the data to the underlying population. During our more recent works, we realized that merging datasets also brings to light issues of time scales. For example, a light sensor can have a sampling frequency of once every fifteen minutes, assuming that the data being tracked relates to the sun's movement. These types of sensors would be challenging to coalesce if the behavior observed is at a smaller time scale, such as once every second. Referencing the Nyquist sampling rate \cite{landau1967sampling}, we suggest aiming for at least twice the sampling rate than the fastest observable instance of the behavior and noting that combining sensor information brings the time scale down to the largest common denominator. Furthermore, having insights that directly relate to a human component (such as occupancy count) instead of a proxy variable (such as CO2) can improve the quality of the observation.

\subsection{Software development is not research}
We found it non-trivial to organize and retain a changing collection of user, device, technical backgrounds, and surveys, and also keep up with the development work necessary to test out new ideas and connect between commercial off the shelf devices and custom-build devices. We realized that we spent a significant amount of time implementing the software stack, of itself does not lead to tangible research outcomes. Furthermore, we realized that there are large variety of team structure for research labs, some of which do not have dedicated software development staff to support the research questions they might have but have great ideas to contribute.

\section{Limitations and Future Works} \label{sec:limits}
We anticipate many interesting research areas to emerge from the use of this system, and also many improvements that can be made with our initial implementation. We list a couple of future directions we anticipate are possible: 

\paragraph{Autonomous Occupancy Polling Stations} Occupant polling stations have been investigated as a strategy to track thermal comfort in a building \cite{lassen2020field}, but the integration of polling stations with robots have yet to our knowledge be fully explored.  By combining the \textit{survey}, \textit{device}, \textit{accounts}, and \textit{floor plans} modules it could be possible to conduct thermal comfort surveys with granular location and time information. 

\paragraph{Simulations and Digital Twins} The current \textit{floor plans} module can be replaced with more sophisticated smart building simulation frameworks to represent larger spatial contexts. For example, Weber et al. demonstrates how caustic light patterns can be predicted reliably using photon mapping for complex 3D-printed glass structures \cite{weber2020photon}. Connecting the system to simulation platforms enable more research into more sophisticated interfaces or machine learning applications, such as tracking user activity through lighting signals, but more importantly this enables researchers to implement the new application into their own living labs.

\section{Conclusion} \label{sec:conclusion}
This paper introduces a bottom-up living lab framework and demonstrates key strategies to implement and maintain operations for a living lab infrastructure. By using the \textit{accounts}, \textit{surveys}, \textit{devices}, \textit{floor plans}, and \textit{dashboard} modules, future researchers are freed to better explore relationships and implementations for the living labs of tomorrow.  

\section{Online Resources}\label{sec:online}
We include at https://github.com/livinglinklab/lll.git the repository containing the source code, and docker for the proposed living lab framework. 

\section{Acknowledgments} \label{sec:ack}
We thank the University of Virginia (UVa) Link Lab and affiliated professors, staff, and students for their contributions and discourse that without which this work would not have been possible. We also want to thank the support of the UVa Facility Management Services team, who supplied us with plans and 3d models of the building.

% \bibliographystyle{ACM-Reference-Format}
\bibliographystyle{unsrtnat}
\bibliography{sample-base}

\appendix
\begin{figure*}
    \centering
    \includegraphics[width=\linewidth]{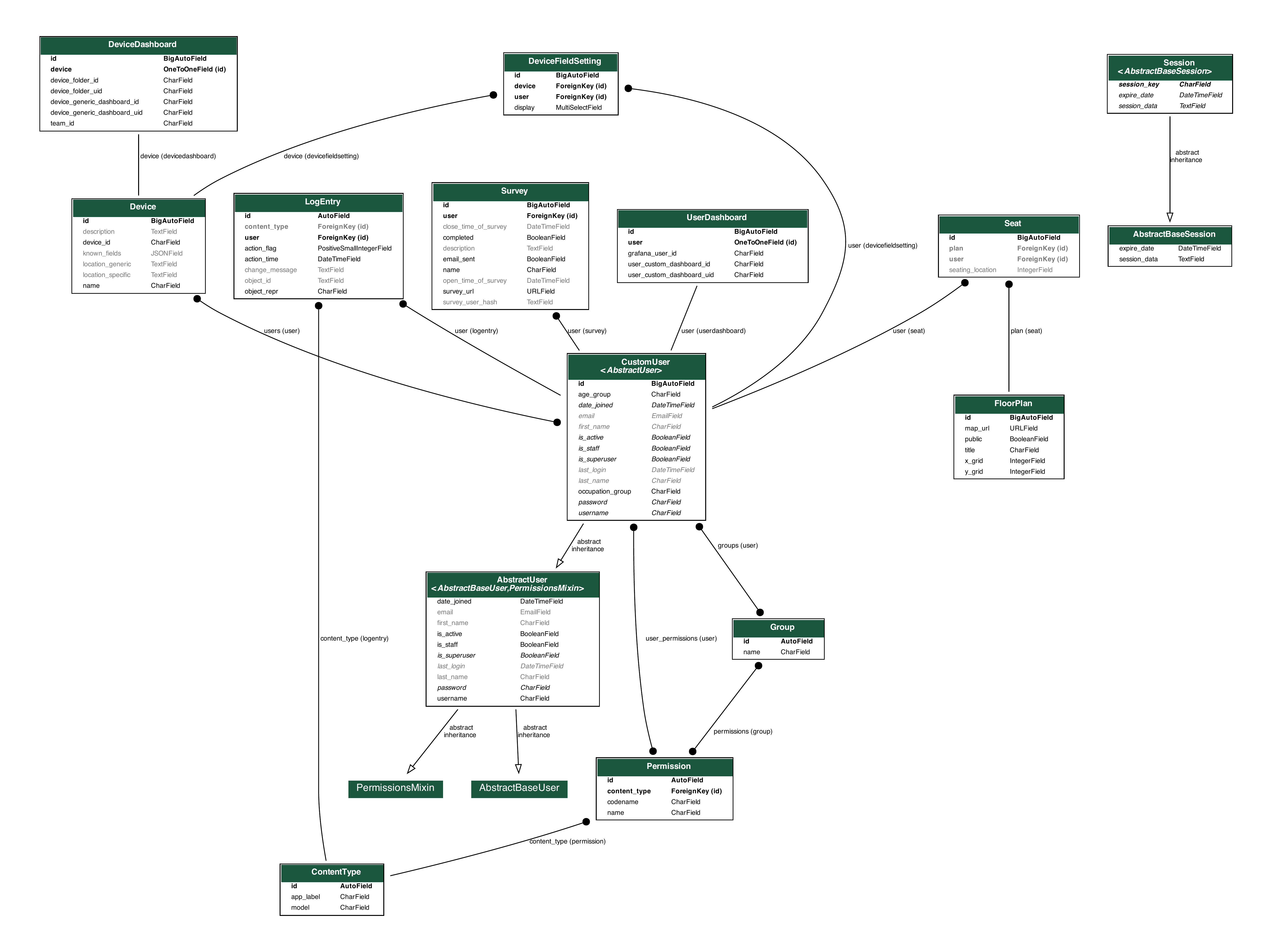}
    \caption{Example model and fields for the proposed living lab infrastructure. By building off of an existing framework, researchers can focus more time developing features directly related to their research question, as opposed to debugging and testing infrastructural connections.}
    \label{fig:lll_model_detailed}
\end{figure*}

\end{document}